\RequirePackage{fix-cm}
\documentclass[smallextended]{svjour3}           
\smartqed  
\usepackage{graphicx}
\usepackage{hyperref}
\usepackage[boxruled, linesnumbered]{algorithm2e}
\usepackage{boldline,multirow,tabularx,colortbl,diagbox,makecell,fancybox}
\usepackage{amsmath}
\usepackage{bbm}
\usepackage{amsfonts,amssymb,amsmath,mathrsfs,array,stmaryrd}
\usepackage{subcaption}
\newcommand{\centered}[1]{\begin{tabular}{l} #1 \end{tabular}}
\begin{document}

\title{Collusion-Resistant Worker Set Selection for Transparent and Verifiable Voting
}

\titlerunning{Collusion-Resistant Worker Set Selection}            

\author{Matthieu Bettinger\footnote{Corresponding author}             \and
            Lucas Barbero               \and
            Omar Hasan
}

\authorrunning{Bettinger et al.} 

\institute{M. Bettinger \at
                  University of Lyon, INSA-Lyon, F-69621, France \\
                  \email{matthieu.bettinger@insa-lyon.fr}               
               \and
               L. Barbero \at
                 University of Lyon, INSA-Lyon, F-69621, France \\ \email{lucas.barbero@insa-lyon.fr}
               \and
               O. Hasan \at
                 University of Lyon, INSA-Lyon, F-69621, France \\ \email{omar.hasan@insa-lyon.fr}
}

\date{Received: 13/06/2021 / Accepted: 21/05/2022}

\maketitle

\emph{“This version of the article has been accepted for publication, after peer review (when applicable) but is not the Version of Record and does not reflect post-acceptance improvements, or any corrections. The Version of Record is available online at:
\emph{https://doi.org/10.1007/s42979-022-01227-5}. Use of this Accepted Version is subject to the publisher’s Accepted Manuscript terms of use https://www.springernature.com/gp/open-research/policies/accepted-\\manuscript-terms”}

\bigskip

\begin{abstract}
Collusion occurs when multiple malicious participants of a distributed protocol work together to sabotage or spy on honest participants. decentralized protocols often rely on a subset of participants called workers for critical operations. Collusion between workers can be particularly harmful to the security of the protocol. We propose two protocols that select a subset of workers from the set of participants such that the probability of the workers colluding together is minimized. Our first solution is a decentralized protocol that randomly selects workers in a verifiable manner without any trusted entities. The second solution is an algorithm that uses a social graph of participants and community detection to select workers that are socially distant in order to reduce the probability of collusion. We present our solutions in the context of a decentralized voting protocol proposed by Schiedermeier et al.~\cite{schiedermeier2019transparent} that guarantees transparency and verifiability. Enabling collusion-resistance in order to ensure democratic voting is clearly of paramount importance thus the voting protocol provides a suitable use case for our solutions.

\end{abstract}

\section{Introduction}
\label{intro}
Voting is an essential element of a robust democracy. However, traditional polling site voting, which requires the voters to physically visit designated locations, poses several problems. For example, in a sanitary context such as the ongoing COVID-19 pandemic, health concerns of in-person voting are a major concern. The requirement of in-person voting may also result in low voter turnout \cite{covid-election-impact}. Mail-in ballots pose their own set of challenges. For example, delays and associated controversies \cite{ballot-deadlines,postal-delay} of the 2020 presidential election in the United States demonstrated the limitations of mail-in ballots. These issues show us that we need a more transparent and verifiable method to ensure democratic votes, which relies less on a central entity or authority. In response to this problem, Schiedermeier et al. \cite{schiedermeier2019transparent}, propose and evaluate a secure electronic referendum protocol for users that ensures confidentiality, integrity, transparency, and verifiability.

In this paper, we use this protocol as an example of a decentralized process that needs to resist against colluding malicious entities. Collusion attacks involve multiple protocol participants working together to sabotage that protocol or steal information from it. We propose solutions for selecting entities in a transparent and decentralized manner that reduce the probability of collusion. Schiedermeier's protocol\cite{schiedermeier2019transparent} indeed operates in a trustless and decentralized network environment, which implies that the participants involved in the protocol do not have to trust each other or any third parties. 
Some steps of the protocol are carried out by a subset of the participants of the referendum called workers, chosen by a trusted entity called the initiator. 
However, we identify that 
the initiator may act maliciously and handpick corrupted workers who can collude together in order to compromise the security of the protocol. This may include discovering the secret votes, corrupting the final result, or preventing a result altogether. We note that the challenge of worker collusion is not specific to the protocol by Schiedermeier et al. \cite{schiedermeier2019transparent}. Worker collusion is a general threat to a range of secure decentralized protocols.

In this paper, we propose two approaches to the selection of a set of workers that minimize the potential of collusion. Firstly, we propose a solution where workers are designated randomly among a set of participants in a transparent, verifiable, and decentralized manner. Secondly, given a social graph of the referendum participants, we use graph analysis and community detection in order to choose workers amongst them such that the social distance between them is maximized and thus the potential for collusion is reduced. We use the protocol by Schiedermeier et al. \cite{schiedermeier2019transparent} as the context for the two proposed approaches in this paper. However, without loss of generality, the approaches may be used in other scenarios where a collusion-resistant subset of workers needs to be chosen from a set of participants.

The outline of the rest of the paper is as follows. In \autoref{initialprotocol}, we introduce the protocol by Schiedermeier et al. \cite{schiedermeier2019transparent} as our exposition use-case. We then discuss previous works related to worker selection, as well as their collusion resistance properties in \autoref{relatedwork}. Next, in \autoref{problemstatement}, we describe and formalize the problem of collusion-resistant worker set selection. Our solutions are detailed in \autoref{Vote order-based} for verifiable random worker selection and \autoref{ComDetWorkerSelection} for verifiable social graph aware worker selection. In \autoref{experimental protocol}, \autoref{dataset_tools}, and \autoref{evaluation}, we focus on the experimental protocol that we use, the dataset and tools used, and the analysis of the experimental results, respectively. This is followed by a discussion of the findings in \autoref{discussion} and the conclusion in \autoref{conclusion}.

\section{Use case -- Transparent and Verifiable Voting}\label{initialprotocol}


In this section, we first describe the transparent and verifiable blockchain-based voting protocol by Schiedermeier et al.~\cite{schiedermeier2019transparent}. We use this protocol as a use-case in this paper for exposition purposes, that is to provide a clear context to support the objectives of subsequent sections: firstly to analyze and compare related works (\autoref{relatedwork}), secondly to highlight risks when selecting workers (\autoref{problemstatement}), and finally to present our proposed solutions (sections \ref{Vote order-based} and \ref{ComDetWorkerSelection}). Nevertheless, our solutions are not limited to Schiedermeier et al.'s protocol\cite{schiedermeier2019transparent}: they can be applied to other contexts with the need for collusion-resistant worker set selection. This section provides a brief overview of the voting protocol, which is sufficient for the purposes of this paper. However, the reader may take a look at the original paper\cite{schiedermeier2019transparent} for complete details.

\subsection{Roles}

The protocol involves the following roles:

\begin{itemize}
  \item  
        \emph{Initiator:} The initiator is an entity that sets up and initiates a referendum. The initiator defines the parameters necessary for the execution of the referendum. This includes the set of voters and the set of workers. The initiation of the  referendum is discussed in further detail in Section \ref{init-phase}.

  \item
        \emph{Voters:} Voters are human users who participate by voting in the referendum. 
        The set of voters participating in an instance of a referendum is given as: $V = \{ v_1, ..., v_k \}$.

  \item
        \emph{Workers:} Workers are users who are a subset of the voters. In addition to voting, a worker's device runs the secure multi-party computation phases of the protocol and contributes to the calculation of the intermediate results of the protocol. The set of workers, which is a subset of the voters, is given as: $W = \{ w_1,...,w_n \}, W \subset V$.
\end{itemize}

\subsection{Ledger}

A blockchain-based ledger $L$ is the unique designated channel of communication during the protocol. It represents an immutable and publicly accessible data container. The messages are exchanged by appending and querying records on this public ledger. Once added in the ledger, a record cannot be updated or deleted. This is due to the immutability and append-only properties provided by a blockchain-based ledger. It also provides transparency due to the public visibility and verification of the blockchain.

\subsection{Protocol Phases}

The protocol is divided into four distinct phases:

\begin{itemize}
  \item{\emph{Initiation:}}
         This phase consists of a single broadcast message $b_{id(init)}$ sent by the initiator to transmit the referendum parameters to all the voters. This ensures that all the participants share the same referendum context.
 
  \item{\emph{Vote submission:}} After a voter $v_i$ retrieves the initiator’s broadcast message from the ledger, he then submits his vote. First, he secretly makes his voting choice which is associated with a specific number value, then he will generate $n$ shares based on this value. Each share is intended for one specific worker $w_j$ and is encrypted by the voter $v_i$ with the $w_j$’s public key. At the end of the phase, the voter persists all the encrypted shares into $n$ messages to the ledger $L$.
 
  \item{\emph{Intermediate result computation:}}
        This phase consists of the computation of intermediate results by the workers. Each worker will retrieve the encrypted shares corresponding to himself from the ledger, decrypt each share with his private key, and contribute to the homomorphic calculation of intermediates results (sum of all votes and sum of all square votes for a later detection of illegal inputs). Finally the worker will persist these results on the ledger $L$.

  \item{\emph{Determination and validation of the outcome:}}
 In this final phase, each voter will retrieve the intermediate result shares. While combining all the results, each voter is able to reconstruct the referendum outcome and the referendum checksum in order to validate these values and detect potential illegal inputs.
\end{itemize}

\subsection{Initiation Phase Details}
\label{init-phase}

The solutions presented in this paper focus on the initiation phase, as this is where the selection of workers is carried out. As described by Schiedermeier et al.~\cite{schiedermeier2019transparent}, the initial broadcast message $b_{id(init)}$ transmitted by the initiator contains the parameters given below. The message is added to the ledger, which allows all participants to retrieve the same referendum parameters. As we will discuss in \autoref{problemstatement}, the selection of the workers by the initiator can lead to collusion issues.

\begin{itemize}
  \item
        The identities of the participating voters: $V = \{id(v_i)|v_i \in V \}$. Public keys of the voters are used as their anonymous identities.
  \item
        The set of workers: $W = \{id(w_j)|w_j \in W\}$.
  \item    
        The referendum question with a binary response. Yes = $+1$, No = $-1$.
  \item
        The subsequent phases of the protocol transition at  specified time-stamps: $q_{1-2}, q_{2-3}, q_{3-4}$. All required input data for a given phase is expected to be submitted to the ledger before the corresponding timestamp.

\end{itemize}

\section{Related work}\label{relatedwork}

We now describe related work dealing with worker selection and collusion-resistance, and discuss to what extent they could be integrated to the voting protocol presented above, or, more generally, to collusion-resistant worker selection use cases. Following categories of works are presented in the order of decreasing needed insight about participants, ending with random selection of workers. The first works are based on monitoring participant behavior and their interactions, while the next two categories use the network topology of participants. Finally, work on random selection of workers is presented.

\subsection{Reputation-Based Witness Selection}
Fighting potential colluding workers through decentralized reputation systems as mentioned in \cite{reputation} could be an interesting idea. There are however two main limitations to this approach given our use-case.

First, participants are pseudo-anonymized for a vote, and that pseudo-identity changes between each voting event. This prevents using past behaviour of participants to choose workers among them, like it is done by Aral et al\cite{aral2020}. In their work, in the context of decentralized cloud computing, they detect patterns of workers that tend to fail together (accidentally or maliciously), in order to choose a worker set that maximizes the probability of success. Because we cannot link pseudo-identities between distinct voting events, we cannot use their approach. Even without pseudo-identities, the rarity of voting events and even rarer selections of a given participant as a worker mean few events when its reputation could be evaluated. This reduces the meaningfulness of computing a reputation score.

Second, an important requirement for any reputation system used in the scenario of voting would be strong privacy preservation. Given a real person, the secrecy of whether that person voted in a given event, and more importantly the content of the vote should be preserved. 
These security concerns regarding the voting phase are discussed by Schiedermeier et al.\cite{schiedermeier2019transparent}, while our work focuses on a transparent and verifiable setup before that vote. That setup is designed to tolerate malicious participants, meaning it does not seek to identify them (e.g. for prosecution purposes). In some cases, it is indeed difficult to differentiate between accidental and malicious actions, for example with crashes. Therefore, as did the original protocol, we do not need to breach participant privacy.

\subsection{Leader Election}
Leader (s)election, for example developed in \cite{leaderElection} and \cite{OptimalLeader}, uses the network topology of participants in the system to select a subset of them as leaders. However, such schemes are not viable solutions for collusion resistance in our specific use case. 
Indeed, the purpose of these works is to choose some nodes as leaders, so as to minimize the distance between each graph node and its closest leader. Rather than distributing voters around leaders (in our case workers), we would prefer to distance workers from one another, which is not the same goal. Moreover, in order to preserve the confidentiality of the vote protocol, the heuristic for such leader selection would focus on the node IDs, which is the only known attribute known about participants. Again, if the initiator is responsible for the input data (i.e. the social graph needed and IDs), it could easily manipulate these IDs to make sure malicious agents are elected.

\subsection{Community detection in graphs}

Instead, given a network of participants, we want to distance chosen workers from one another. We use community detection notably to reduce the complexity of the algorithm in practice (further detailed in \autoref{ComDetWorkerSelection}). 
Community detection partitions a graph based on its structure into one or more communities, in some cases using edge or node attributes (which is not the case in our work). The objective is to form groups of nodes that are well connected with each other, which form a community, while these nodes are less connected to other nodes outside their community. 

Many algorithms have been proposed to perform this task. In our work, we used the Python igraph\cite{igraph} library for computations on graphs, which proposes most of the recent and leading algorithms in community detection. This library notably includes:
\begin{itemize}
        \item Rosvall and Bergstrom's InfoMap\cite{Rosvall2008}. It performs a random walk on the edges of a given graph, based on the intuition that a random walker is likely to stay in a well-connected subgraph (i.e. a community), once it has entered it;
        \item Clauset et al's Fast greedy modularity optimization\cite{clauset2004}. It optimizes the Newman and Girvan modularity metric\cite{Newman2004Finding}, which compares the structure of a partitioned graph with a graph generated randomly.
        \item Blondel et al.'s multilevel algorithm\cite{Blondel_2008}. As described by Lancichinetti and Fortunato\cite{Lancichinetti2009}, it locally optimizes modularity, such that it yields a better estimate than Clauset et al.'s\cite{clauset2004}, while running on a near-linear complexity on the number of edges.
\end{itemize}

For the purposes of our experiments, we used Blondel et al's multilevel algorithm\cite{Blondel_2008}, because it is deterministic and because of its properties presented before.
Moreover, based on Lancichinetti and Fortunato's analyses\cite{Lancichinetti2009}, it also performs well compared to other approaches. Subsection \ref{ComDetAlgo} goes in further detail as to why we chose the multilevel algorithm for our experiments.

\subsection{Decentralized random number generation and worker selection}

Our first proposed solution relies on random selection of a set of workers among participants. We use a blockchain as an immutable and transparent messaging hub for participants, so they can generate a random seed number in a decentralized manner in a single phase, which determines the set of workers.

A work close to our problem is Nguyen-Van et al.'s\cite{ginarPaper}, who propose a decentralized multi-step protocol based on Homomorphic Encryption, Verifiable Random Functions (VRF) and distributed ledgers (e.g. blockchains) to generate random numbers. Homomorphic encryption enables adding (or multiplying) values while they are encrypted. VRFs generate random numbers as well as proofs that these numbers were obtained by running that VRF. The initiator in Schiedermeier et al.'s work\cite{schiedermeier2019transparent} could request a random number to be generated and would also obtain a proof of correct execution through the pipeline. However, a limitation is that the initiator must not share the private key it uses for requesting the number, otherwise voiding tamper-resistance on the result. As the initiator is not trusted in our case, this method cannot be used.

Alternatively, their protocol includes a step where each candidate in the random number generation runs a VRF: if the returned value is higher than a given threshold, they will indeed participate in generating the number. This is in fact a random worker set selection. The difference with our approach is that their set size follows a binomial distribution, whereas in our solution the size is predetermined. A deterministic size may be preferable in our use-case, so the vote does not run a probabilistic risk of having too few workers to be resistant to attacks.

Simic et al.\cite{Simic2020ReviewRNG} reviewed mechanisms, including those presented above, for generating random numbers in decentralized environments, in particular in blockchains.

\section{Problem Statement} \label{problemstatement}

\subsection{Adversary Model}

We will call malicious all potential workers as well as the initiator likely to work together in order to form a coalition with the intention of disrupting the expected behavior of the referendum. We can distinguish three malicious behaviors resulting from workers' collusion:

\begin{itemize}

        \item Discovery of the secret votes of the honest participants from the intermediate values.
        \item Manipulation of intermediate values in order to corrupt the final result.
        \item Prevention of the computation of the final result due to inactivity from the malicious workers.

\end{itemize}

The initial list of participants in the voting event is trusted. Notably, each pseudo-identity in the list corresponds to a unique and real participant, for example an officially registered citizen. This means that a malicious entity cannot create multiple fake pseudo-identities to sway the vote, namely, Sybil attacks are not possible. Similarly, people expected to participate in the vote are present in the list. We accept both assumptions for the following reasons:

\begin{enumerate}
        \item An expected participant whose pseudo-identity is missing in the list can easily detect and report it, as the list is publicly available;
        \item Reason (1) also means that for a list size equal to the expected number of real participants, malicious entities cannot replace the identities of other participants by theirs;
        \item A list containing more pseudo-identities than expected real participants is likewise detectable and shows the presence of malicious fake participants, given reasons (1) and (2);
\end{enumerate}

Note however that this does not prevent some other actions by malicious entities that increase their power in the vote, like:
\begin{itemize}
        \item bribing a real participant;
        \item obtaining control over a pseudo-identity (e.g. through phishing). This non-consensual loss of control by the honest participant may be reported and fixed before the voting event takes place, canceling the malicious action.
\end{itemize}
The effect of such actions is an increased number of malicious entities in the protocol. In this work, we simply count $m$ maliciously controlled pseudo-identities in the participant list during an execution of the protocol.




\subsection{Collusion-Resistant Worker Selection}

An arbitrary selection of workers by a single entity (the initiator in the case of Schiedermeier et al.'s protocol\cite{schiedermeier2019transparent}) can lead to collusion if the entity performing the selection is malicious. 
This is why we propose to remove this arbitrary choice, replacing it with a more neutral and secure process.

We will now define which constraints need to be upheld by our solutions to successfully achieve this task. Let $m$ be the number of malicious workers among all $n$ workers. 
We place ourselves in the context of a $t-n$ threshold-based Shamir secret sharing scheme~\cite{ShamirSecret}, which allows splitting up a secret into $n$ shares (each held by one of the $n$ workers) in such a way that any $t$ of them suffice to reconstruct the secret. Malicious agents cannot gain any information about the secret if they possess strictly less than $t$ shares.
\begin{enumerate}
        \item
        To ensure the secrecy through Shamir’s threshold system-based secret sharing scheme~\cite{ShamirSecret}, the number of malicious workers $m$ shall not be more than $t$;
        \item
        If there are less than $t$ honest active workers, then a coalition of malicious workers can prevent reconstructing the value by being inactive. Therefore, $m$ shall not surpass $n-t$ workers. Like the first constraint, this rule considers the presence of $m$ malicious agents in a $t-n$ threshold system;
        \item A checksum is computed at the same time as the referendum's result. Its purpose is to count the total number of votes. This prevents participants from voting illegal referendum values. For the referendum’s checksum reconstruction, $m$ must be less than $n-t^2$ workers (again against inactivity)~\cite{10.5555/36664.36683}.
\end{enumerate}
Therefore we have:

        \begin{subequations}
        \begin{align}
            n\geq t>m\geq0 \label{eqConstr1}\\
            n-t>m\geq0 \label{eqConstr2}\\
            n-t^2>m\geq0 \label{eqConstr3}
        \end{align}
        \end{subequations}

The collusion limit is when $m=t-1$, i.e., when the malicious coalition is missing one worker to be able to discover the secret.  Replacing $m=t-1$ into (3) gives us: $t^2+t-1-n<0$ and $t>0$. Given $n$ workers, the maximal $t$ is therefore given by $t_{max}=\lfloor(-1+\sqrt{5+4n})/2\rfloor$, based on verifiability constraints presented above.

\begin{figure}
 \includegraphics[width=\linewidth, keepaspectratio]{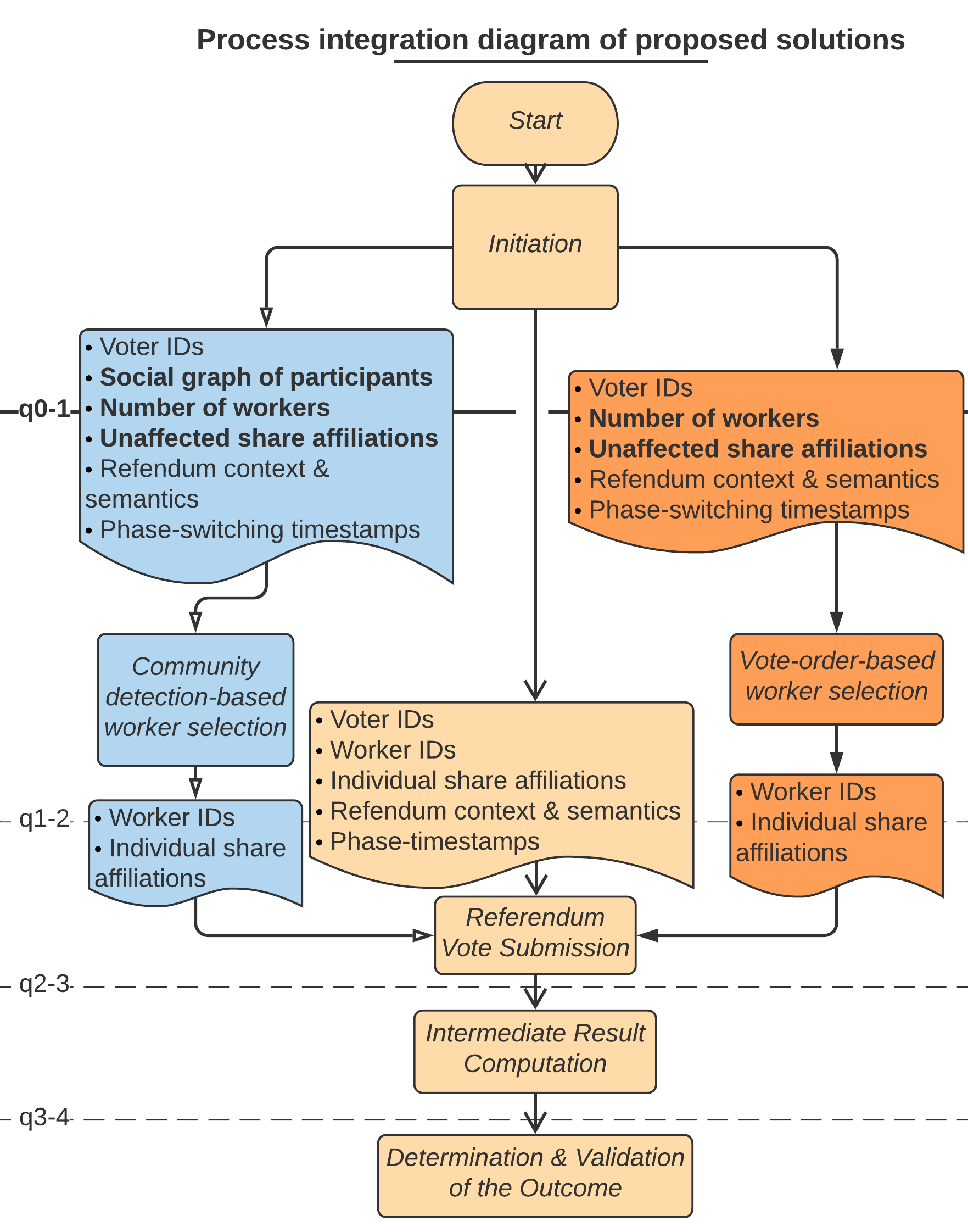}
 \caption{Process diagram of the initial protocol (in the center), with proposed solution protocols integrated onto it on the sides. $q_{i-j}$ lines represent phase transitions as defined in \cite{schiedermeier2019transparent}, $q_{0-1}$ is a new transition used only in vote-order-based worker selection (right-hand side). Rectangles represent process phases, ``document''-shapes represent messages on the ledger. Document-shapes contain data items used in the protocol. Bold items are changes compared to the initial protocol.
 }
 \label{fig:ProposedSolutions}
\end{figure}




\section{Proposal 1: Verifiable Random Worker Selection} \label{Vote order-based}

Replacing arbitrary worker selection with a verifiable random selection enables quantifying the probability of collusion. For example, in \autoref{evaluation}, we show that even with a third of the participants as workers in the sampled social graph, the probability of collusion is less than 2.5\%. Before, that probability was unknown, because the initiator was a trusted entity. Now, however, the probability $\mathcal{P}(X=m)$ of having sampled $m$ malicious workers among $M$ malicious participants knowing $n$ total workers were sampled among $P$ total participants follows a hypergeometric law $\mathcal{H}(n,\frac{M}{P},P)$. This probability is given by:
\begin{equation}
        \mathcal{P}(X=m)=\frac{\binom{M}{m}\binom{P-M}{n-m}}{\binom{P}{n}}
\end{equation}
For example, let us assume that we want to have a probability of collusion of less than 5\% and expect to have $M$ malicious participants out of the total $P$ participants. Plotting the hypergeometric law's Cumulative Distribution Function $CDF(n,m)$, for given values of $M$ and $P$, and taking the intersection with the plane of probability 95\%, gives us the curve $\mathcal{C}(n)$ such that there is 95\% probability that there are less or equal than $m$ malicious workers in a sample of $n$ participants.

Our protocol computes a random number in a decentralized fashion. This number is used as a seed to randomly select workers among participants. This seed will be known by all. Therefore, the whole procedure can be verified by any participant. This solution introduces an additional phase in the voting protocol. This phase harnesses the immutability and ordering properties of messages committed on the ledger to generate a verifiable seed. Indeed, we can use these properties to generate a number in a decentralized manner by soliciting the participation of voters, who will deposit a message on the ledger. Our number is generated by comparing the order in which a subset of participants committed a message on the ledger, as well as the composition of that subset, to the ordered complete list of participants. Given the number generation algorithm and the messages on the ledger, any participant can verify the outcome of that phase.

The public ledger on the blockchain has the particularity to be immutable. It means that malicious entities cannot interfere with the ordering of messages on the ledger nor can they interfere with whether messages are published or not on the ledger. 
Blocks are appended sequentially to the blockchain, and each block contains a list of transactions (in our case our messages). We can use this sequential data structure to define an order of participant IDs.
\subsection{Context}

\paragraph{Underlying blockchain characteristics:}

Our solution relies on blockchain architectures that meet the following specifications:
\begin{itemize}
        \item Consensus algorithm: should be so the malicious coalition lacks the necessary resources to manipulate or control it. For example, it can be based on hardware resources, like Proof-of-Work, or cryptocurrency assets, like Proof-of-Stake or its Delegated Proof-of-Stake variant.

        \item Openness/Access control: the blockchain system may either be public or permissioned. However, public blockchains are useful to have a large number of nodes maintaining them, which enables high decentralization of the consensus algorithm. The voting application logic would run on top of this consensus layer
        . Indeed, participants in the vote can be registered and identified in the vote's participant list by their public blockchain account addresses, or their corresponding public keys (for example with Ethereum\cite{ethereum}, the account address is derived from a user's public key). Participants are therefore registered at the application-level (or smart contract-level, if present), while the underlying blockchain is open to the general public. This mechanism makes it easy to distinguish real from fake vote participants, in the context where the voting application coexists with other services (and their respective users) on the same blockchain. Because of the extended blockchain user base, which includes nodes maintaining the network, it also means participants are only required to be able to send transactions (containing the protocol messages) to the blockchain using their account. In turn, this means participants do not need to run a full blockchain node in order to participate in the voting protocol. For example, with Ethereum\cite{ethereum}, light clients enable to verify the validity of blocks and transactions, without requiring the storage of the whole blockchain historic data.

        \item Smart contracts: the protocol only needs the blockchain as a messaging hub. As such, advanced features like smart contracts are not required. Nonetheless, smart contracts are useful to enable decentralized and easier enforcement of the protocol (e.g. for phase-switches, for considering only a voter's latest vote submission).
\end{itemize}

\paragraph{Inputs:}

Let $\mathcal{P}$ be an ordered list of $P$ participant IDs, obtained through a trusted process (e.g. voter registration by government entities). Let $n$ be the size of the subset of participants to select in $\mathcal{P}$.
Furthermore, let $\mathcal{M} \subset \mathcal{P}$ be the set of malicious participants. Malicious participants are \emph{a priori} indistinguishable from honest participants, i.e. no ID in $\mathcal{P}$ can be proven to be in $\mathcal{M}$ before the protocol starts. $M=|\mathcal{M}|\geq 0$ the number of malicious participants is unknown as well.

\paragraph{Outputs:}

Let $\mathcal{W} \subset \mathcal{P}$ be the set of selected workers.

\paragraph{Objective:}
Select $n=|\mathcal{W}|$ workers out of $\mathcal{P}$, such that the procedure is verifiable by all participants in $\mathcal{P}$ and all malicious participants in $\mathcal{M}$ cannot increase their chances of being chosen as workers to more than random chance (i.e. the probability of $m$ malicious participants being selected as workers follows a hypergeometric law $\mathcal{H}(n,\frac{M}{P},P)$).

\paragraph{Hypotheses:}
\begin{itemize}
  \item[($h_{1}$)] Malicious entities cannot interfere with the block-/transaction-ordering on the ledger. \emph{Rationale: Property guaranteed by the Blockchain};
  \item[$(h_{2}$)] Malicious entities cannot interfere with whether transactions are published or not on the ledger. \emph{Rationale: It is assumed that the coalition of malicious entities does not have the resources to take control over the blockchain's consensus algorithm (as required by blockchain characteristics defined before)};
  \item[$(h_{3}$)] If a Cryptographically Secure Pseudo-Random-Number Generator\footnote{Cryptographically Secure Pseudo-Random-Number Generator: high entropy generator which resists reverse-engineering, cryptanalysis and passes statistical randomness tests} (CSPRNG), a high entropy generator, and a numbering method with a large output space are used to generate the list of workers, malicious participants cannot increase their chances of being chosen as workers by choosing particular IDs. \emph{Rationale: Property guaranteed by CSPRNGs.}
\end{itemize}
\subsection{Algorithm}

A new phase is added between the referendum’s initiation and the referendum vote. Initiation commits a list of $P$ participants and a number $n$ of workers to select. This new phase will consist of the participants choosing to deposit a message on the ledger or not. This creates a sublist of participant IDs on the ledger which will be used to generate a seed number for a CSPRNG. This CSPRNG will in turn sample $n$ workers among the $P$ participants. It is important to note that these committed messages are distinct from the participants' actual votes, which happen in a later phase.

This protocol proceeds as follows (see \autoref{fig:ProposedSolutions} above for reference):
\begin{enumerate}
        \item The phase begins when the phase-switching deadline $q_{0-1}$ after the referendum’s initiation is reached. This happens when the initiator commits the referendum's information on the ledger. These information include:
        \begin{itemize}
            \item IDs of participants able to vote in the referendum;
            \item Number $n_{workers}$ of workers to select;
            \item Unaffected share affiliations between participants and workers used in Shamir's Secret Sharing Scheme (i.e. a participant $x$ will divide its secret referendum-vote in the next phase and respectively give each share to the $k^{th}, l^{th},...$ selected workers);
            \item Context and semantics for the referendum;
            \item Phase-switching timestamps $\{q_{1-2}$, $q_{2-3}$, $q_{3-4}\}$.
        \end{itemize}
        \item Each participant decides to commit a message on the ledger or not;
        \item The phase ends when the phase-switching deadline $q_{1-2}$ is reached;
        \item An ordered list of messages $V$ will then be present on the ledger. A number is computed from $V$ and the ordered list of all referendum participants $P$. Three main computations can be used to generate this number: the permutation, combination and arrangement numbers of $V$ in regard to $P$. The first computation uses only the message-list $V$ as randomness input, while both $V$ and $P$ are used for combination and arrangement number generation. Mathematical expressions and output spaces are presented in \autoref{tab:Numbering}.
        \begin{itemize}
        \item The computed number is used as a seed for a CSPRNG. Then, $w$ workers are sampled in the list of participants with this generator;
        \item Shamir's Secret Sharing Scheme's shares affiliations between participants and workers for the referendum-vote are then attributed to each selected worker;
        \item The protocol can then proceed to the next phase (referendum vote).
\end{itemize}

\end{enumerate}

\paragraph{Remarks:}
\begin{itemize}
        \item \emph{On multiple committed messages:} only the first message deposited by each participant (if they deposited any) is considered. This prevents seed value manipulations from malicious participants as the phase goes on.

        \item \emph{On merging worker selection with the voting phase:} as said earlier, the workers are selected during a distinct phase preceding the actual vote. For simplicity, and given worker selection is based on an order of messages, one would intuitively want to perform worker selection during the vote itself. However, this is impossible due to an incompatibility with the voting protocol mechanisms as proposed by Schiedermeier et al.\cite{schiedermeier2019transparent}. Indeed, participants vote by committing $n$ distinct shares, encrypted with each worker's public key (so that only that worker can decrypt it with its private key). This means that all workers must be identified prior to the voting phase. Moreover, from a security standpoint, merging the phases reduces the entropy of the number generation, because its entropy comes from both the number of messages and their order. Indeed, honest participants are likely to actually vote, which ultimately reduces the space from which the number is generated.
    
        \item \emph{In the absence of committed messages:} a number is generated however many participants commit a message, in particular even when the ordered list of committed messages is empty. In that scenario, using the proposed numbering functions (see \autoref{tab:Numbering}), the obtained value would be zero. The absence of commits is therefore not problematic as long as the hypothesis $(h_{2})$ defined above holds. If $(h_{2}$) is compromised, then the participant list, trusted by the protocol, could be maliciously designed such that no commits (or only malicious commits) make it so enough malicious workers are selected. This would be done by giving malicious participants specific IDs.
    
        \item \emph{Message list traversal:} In the expressions of \autoref{tab:Numbering}, the list of committed messages is traversed in the anti-chronological order (i.e. the latest message corresponds to index $i=0$) for additional security. This prevents malicious participants from progressively computing the updates of the seed value as messages are committed. Indeed, using that direction for list traversal, all indices point to a different value each time a new message is committed.
\end{itemize}

\begin{table*}[ht]
        \caption{Three methods to attribute a number to a list V of comparable non-reoccurring elements (integers, character strings,...) knowing its superlist P.}
        \centering
        \begin{tabular}{lcc}
            \hline\noalign{\smallskip}
            \emph{Numbering Method} & \emph{Expression} & \emph{Output Space} \\
            \noalign{\smallskip}\hline\noalign{\smallskip}
            
            \centered{Permutation \\ $P\!N:V\rightarrow \mathbb{N}$} &
            \centered{$\sum_{i=0}^{|V|-1}(i!\sum_{k=0}^{i-1}\mathbbm{1}_{x_{k}<x_{i}})$} &
            \centered{$\llbracket 0;|V|!\llbracket$}\\
            
            \noalign{\smallskip}\hline\noalign{\smallskip}
            \centered{Combination \\ $C\!N:V,P\rightarrow \mathbb{N}$} &
            \centered{$\sum_{k=0}^{j(0)-1}\binom{|P|-j(0)+k}{|V|-1}$ \\$+ \sum_{i=1}^{|V|-1}\sum_{k=0}^{j(i)-j(i-1)-2}\binom{|P|-j(i)+k}{|V|-i-1}$}&
            \centered{$\llbracket 0;\binom{|P|}{|V|}\llbracket$}\\
            \noalign{\smallskip}\hline\noalign{\smallskip}
            \centered{Arrangement\\$A\!N:V,P\rightarrow \mathbb{N}$} &
            \centered{$\sum_{i=0}^{|V|-1}(\frac{|P|!}{(|P|-i)!}+C\!N(V,P)*|V|!+P\!N(V))$}&
            \centered{$\llbracket 0;\sum_{v=0}^{|P|} A_{|P|}^{v}\llbracket$}\\
            \noalign{\smallskip}\hline
        \end{tabular}
        \\With: $\mathbbm{1}_{F}=\begin{cases}
                0, F$ is False$\\
                1, F$ is True$ \\
                \end{cases}$, $F$ a boolean expression. $P=[x_0,..,x_{p-1}]$ a list of elements in ascending order, $p$ the number of participants. $x_k$ is the ID of the $k$-th participant in the referendum. $V=[y_0,..,y_{v-1}]$ ordered by the order of messages in the ledger, $V$ a sublist of $P$, $v$ the number of messages. $j:\mathbbm{N}\rightarrow \mathbbm{N}$ a function such that: $j(i)=k$ iff $y_i=x_k$, i.e. for an ID in $V$ of index $i$, returns its index in $P$. $A_n^k=\frac{n!}{(n-k)!}$ is the amount of distinct arrangements of $k$ among $n$ non-reoccurring elements, also called $k$-permutations of $n$.
        \label{tab:Numbering}
\end{table*}

\subsection{Discussion}\label{Vote order-based properties}

Malicious entities cannot interfere with the presence and order of honest message deposits ($h_{1}$, $h_{2}$). Therefore, a favorable condition for them is to wait until all honest participants have most likely made their choice whether to deposit a message on the ledger. From the malicious perspective, these honest messages constitute a sort of nonce for the seed number. If the initiator is malicious, it might also try to give specific IDs to some malicious participants, in order to have them chosen. Note that in the case of Schiedermeier's protocol\cite{schiedermeier2019transparent}, IDs are public keys. This renders it complicated to generate a key-pair such that it favors malicious participants. Moreover, ID assignment happens before this protocol's execution, therefore it is only possible to try to maximize chances of being selected given a set of nonces. The greater the numbering function’s output space, the lower these chances. Enabling participants to cancel their previous message by committing a new one gives more freedom for malicious entities to adapt to the order of honest participant messages. The security of the seed generation is given by the size of the output space, meaning the possible entropy of the number generation, coupled with mechanisms to hinder malicious behavior adapting as the protocol advances.

Our seed generation method's output space depends on the number of participants ($\llbracket 0;\sum_{v=0}^{|P|} A_{|P|}^{v}\llbracket$). Therefore, for low participant numbers, entropy might be too low to guarantee pseudo-randomness. A solution to this problem would be to introduce entropy into the system, by considering a value posted in each participant's message. However, this approach introduces possibilities for malicious entities to manipulate the seed value.
One must keep in mind that the size of the output space is a sum of factorials depending on the number of participants: for 10 participants, there are $10^{6}$ possibilities, but $10^{158}$ for 100 participants. To boost the number of possibilities for a low number of participants, multiple message commits per participant can be handled for number generation. If $n$ message commits are allowed per participant, then each message (containing a number between 1 and $n$) would be considered as that of an artificial participant in a list of $P'=n*P$ participants. Note that the actual proportion of malicious participants (real and artificial) remains unchanged, but the output space for generating numbers greatly increases with $P'$ (see the $AN$ function's output space in \autoref{tab:Numbering}).


\section{Proposal 2: Verifiable Social-Graph-Aware Worker Selection} \label{ComDetWorkerSelection}

Instead of randomly choosing the workers among the voters, this next solution uses social graph data in order to select them. This process is done with a community detection algorithm. First, we compute the communities of the graph. We then decide how many workers will be selected from each community. Finally, we select the workers from the communities whilst trying to maximize their distances to other workers.


If we are able to distinguish several communities in a social graph representing interactions between the candidates of the protocol, we can gather new information to achieve worker selection.
Indeed, two participants close to each other in the same community are likely to know each other and have social interactions\cite{Hasan2011}, in other words, should they also be malicious, they are likely to collude. This assertion leads us to a new algorithm: maximizing distances between workers inside each community and in the whole graph to minimize interactions.

We assume that the participants have access to an anonymized social graph of the participants. This is a simplification of the context for experimental purposes, where privacy of participants is not required. Indeed, a main purpose of this solution is to compare our random worker selection to a selection taking advantage of the social structure of participants. If this graph is available and the number of workers to be selected is known, then this protocol is verifiable by all participants. 
If such a graph is not available, we propose to use our first solution (\autoref{Vote order-based}), which is fully decentralized. In \autoref{discussion}, we also discuss some potential alternatives to the centralized availability of the social graph.

In any case, like our first solution, this approach is independent from the voting protocol that we use to contextualize them. Therefore, it can be useful in other contexts, where such a graph is available and privacy is not a concern. For example, in the context of a referendum in a company, the organization chart, which is open and not privacy-sensitive information inside the company, can be used as an approximation of the interactions between employees, and as such, as an approximation of the social graph inside the company. In this specific context, an ability to punish detected malicious activities may be wanted. Otherwise, the organization chart may be modified to reduce information identifying specific employees. Note that it is possible to enable linking workers to real identities, while preserving voter secrecy, by maintaining two lists of participant identities, one for voting and one for (potentially) working. Geographical clustering of voters, such that a sufficient k-anonymity is guaranteed (for example, with the U.S.A.: federal, state, county or municipality levels), could also be used to approximate the overall social graph from which workers can be selected.

\subsection{Context}

Consider the existence of a ``real'' social graph representing the relationships between vote participants in real life. This abstract structure is to be distinguished with graphs obtained from social networks like Facebook, which are created based on user interactions on that service. As such, the latter are an approximated representation of the complete and theoretical ``real'' graph, with missing or added nodes and edges. Nonetheless, experimentally, we will tolerate this approximation and use Facebook social graphs as datasets (see \autoref{dataset_tools}). In our context of collusion-resistance, the proximity between nodes is considered to represent their likeliness of complicity, meaning workers are more likely to collude together if they are close in the social graph. Therefore, in that context, selecting workers such that they are distant from each other in the social graph would reduce the probability of collusion.

\paragraph{Inputs:}
Let $G$ be an undirected unweighted social graph with $G=(V, E)$. The same reasoning can be applied to a weighted graph. $V$ a set of vertices representing each referendum participant. Each vertex $v$ possesses an attribute which is the participant’s ID. Let $E$ be a set of edges, two vertices are connected with an edge $e$ if the corresponding participants mutually and directly know each other. 
Let $n_{workers}$ be the number of workers which shall be designated among the participants.

\paragraph{Output:}
Let $W$ be a set of participant IDs which shall act as workers in subsequent phases.

\paragraph{Objective:}
Choose $W$ such that workers don’t know each other well enough to form a coalition capable of invalidating the results of the referendum. This means determining $W$ a subset of $V$, such that $W$ contains less than $t$ mutually cooperating malicious nodes out of $n_{workers}$ nodes, i.e.  maximize the social distance between all workers.

\paragraph{Hypotheses:}
Concerning community detection, we consider the following hypotheses:
\begin{itemize}

        \item Two participants know each other less the longer the shortest path between their respective vertices is. This hypothesis is based on heuristics used in topology methods for link prediction problems in (social) graphs, e.g. common neighbors\cite{Hasan2011};

        \item Participants from distinct communities know each other less, even more so if those communities don’t communicate directly with each other (paths between members of the two communities go through other communities).
\end{itemize}



\subsection{Algorithm}
\begin{algorithm}
 \KwData{$G$: unoriented graph $(V,E)$, with $V$ a set of vertices representing each participant, $E$ a set of edges such that $v_i$ and $v_j$ are adjacent if participants i and j know each other; $n_{w}$: amount of workers to place in $G$}
 \KwResult{$W$: subset of vertices $W\subseteq V$, the set of chosen workers}
 $P, C \leftarrow detectCommunities(G)$\;
 \tcc{$P$ partitions $G$ in $n_c$ clusters and attributes a cluster ID $c_k \in V_C=\{i\in \llbracket1;n_c\rrbracket | c_i\}$ to each vertex in $V$, such that $P:P(v \in V)=c_k \in V_C$}
 \tcc{$C=(V_C,E_C)$ an undirected graph of $n_c$ communities. $c_i$ and $c_j$ are connected with an edge $e_{c_{i,j}}\in E_C$ if one or more vertices in $\{v\in V | P(v)=c_i\}$ are adjacent in $G$ to one or more vertices in $\{v\in V | P(v)=c_j\}$.}
  $N \leftarrow quantifyWorkersPerCommunity(G,P,C,n_{w})$\;
  \tcc{$N=\{i\in \llbracket 1;n_{c}\rrbracket | n_i=$ number of workers in $ c_i \}$}
  $W \leftarrow$ empty set\;
  \For{$c_i\in V_C$}{
   $W_i\leftarrow selectWorkers(G,P,c_i,n_i)$\;
   $W \leftarrow W\cup W_i$\;
   }
   \Return{$W$}\;
 \caption{Social-graph-aware worker selection}
  \label{algo:AssignWorkers}
\end{algorithm}

Our main Algorithm \ref{algo:AssignWorkers} can be divided in three steps:
\begin{enumerate}
        \item{\emph{Community Detection:}} From a social graph, we apply a clustering algorithm to find groups of nodes (see \autoref{ComDetAlgo});
        \item{\emph{Quantifying the number of workers to be selected in each community:}} Depending on the number of communities $n_{communities}$ and total number of workers $n_{workers}$, we will compute the number of workers $n_i$ to be selected in each community $c_i$ (see Algorithm \ref{algo:DistribWorkers}):
        \begin{itemize}
            \item[$\rightarrow$] $n_{communities}=n_{workers}$: one worker per community;
            \item[$\rightarrow$] $n_{communities}>n_{workers}$: one worker per community in distant communities;
            \item[$\rightarrow$] $n_{communities}<n_{workers}$: distribute multiple workers amongst communities using a criterion.
        \end{itemize}
        \item{\emph{Selection of workers:}} According to the number of worker $n_i$, we will then choose which participants will become workers for each community and maximize their distances to each other (see Algorithm \ref{algo:SpreadNodes}).
\end{enumerate}
 
\subsection{Step 1: Community Detection}\label{ComDetAlgo}

The choice of the community detection method will greatly impact the worker selection algorithm and the collusion resistance. Indeed, our main hypotheses for workers selection and assignment are based on inter- and intra-communities interactions. We distinguished several criteria to choose our clustering algorithm:
\begin{itemize}
        \item deterministic outcome (no random operations to generate clusters) required for reproducibility and verifiability by participants;
        \item linear or near-linear time-complexity on the number of vertices or edges;
        \item each vertex is assigned to a unique cluster (i.e. no multi-community detection);
        \item the number of detected communities depends only on the graph structure (i.e. no need of parameters to set or vary the number of communities found).
\end{itemize}

Considering the above criteria as a filter for algorithms already implemented in the graph library that we use (igraph\cite{igraph}), we retain the multi-level community detection by Blondel et al.\cite{Blondel_2008} as our community detection algorithm.
\subsection{Step 2: Quantifying the Number of Workers to be Selected in Each Community}

This step consists of determining the number of workers which shall be selected in each community in Step 3 (\autoref{WorkerSelection}), considering the total number of workers $n_{workers}$, the number of communities $n_c$, and a distribution criterion.

\subsubsection{Hypotheses}
We consider the following hypotheses:
\begin{enumerate}
        \item Given a shortest path length $d$ in the graph, two participants in the same community distanced by $d$ know each other more than two participants from distinct communities with that same distance $d$ between them.
        \item Sparse graphs are better suited to house more workers than dense one. \emph{Rationale: there are more shortest paths with values of 2 or more, meaning participants of that same cluster know each other in varying degrees of separation.}
        \item  Placing workers evenly across the graph (i.e. all workers distant by 3 more edges versus creating small clusters of less than $t$ workers) gives better guarantees for non-collusion. \emph{Rationale: each worker does not know its nearest worker-neighbors that well.}
\end{enumerate}

\subsubsection{Algorithm}
If the ratio $\frac{n_{workers}}{n_c}$ is less than one, then workers should be placed in communities distant from each other.
Otherwise, all communities shall contain a worker before assigning multiple workers to a community (to take advantage of hypothesis 1). If there are more workers than communities, the graph structure should be used to infer how many workers should be contained in each community (see algorithm \ref{algo:DistribWorkers}). In this last case, the number of workers in each community is given by the pro-rata of the number of vertices in a community to the total number of vertices in the graph, e.g. a community comprising 10\% of the vertices will receive 10\% of the workers.

\begin{algorithm}
 \KwData{$G$, $P$, $C$, $n_{w}$ as defined in algorithm \ref{algo:AssignWorkers}}
 \KwResult{$N=\{i\in \llbracket 1;n_{c}\rrbracket | n_i=$ number of workers in $ c_i\in C \}$, $n_c$ the number of communities in $C$}
        $N\leftarrow\{i\in \llbracket 1;n_{c}\rrbracket | n_i=0 \}$\;
        \uIf{$n_c=n_{workers}$}{
            $N\leftarrow\{i\in \llbracket 1;n_{c}\rrbracket | n_i=1 \}$\;
        }
        \uElseIf{$n_c>n_{w}$}{
            $C_w \leftarrow spreadVertices(C, n_{w}$)\tcc*{Algorithm \ref{algo:SpreadNodes}, $C_w\subset C$}
            
            $N\leftarrow\{i\in \llbracket 1;n_{c}\rrbracket | n_i \begin{cases}
                0, c_i\notin C_w\\
                1, c_i\in C_w \\
                \end{cases} \}$\;
        }
        \Else{
            $N\leftarrow criterionDistrib(G,P,C,n_{w})$\;
        }
   \Return{$N$}\;
 \caption{Quantifying the number of workers to be selected in each community}
 \label{algo:DistribWorkers}
\end{algorithm}

\subsection{Step 3: Worker Selection}\label{WorkerSelection}
\begin{algorithm}
 \KwData{$G$: unoriented graph $(V,E)$, $V$ the set of vertices, $E$ the set of edges; $n$: amount of vertices in $V$ to select; $V_{av}$: $V_{av}\subset V$, $V_{av}\leq|V|-n$, a set of vertices to avoid (empty by default)}
 \KwResult{$V_S\subseteq V, |V_S|=n$, the set of selected vertices}
        $n_r\leftarrow n$\tcc*{remaining vertices to select}
        \eIf{$|V_{av}|=0$}{
            $V_S\leftarrow$ \{selectInitialVertex(G)\}\;
        }
        {
            $V_S\leftarrow V_{av}$\;
            $n_r\leftarrow n_r-1$\;
        }
        $M_{sp}\leftarrow B\!F\!S(V_S, V)$\tcc*{Breadth-first-search distances from vertices in $V_S$ to vertices in $V$ into a $(|V_S|,|V|)$ matrix. If the graph is weighted, then BFS can be replaced with an adequate algorithm.}
        \For{$i \in \llbracket1;n_r\rrbracket$}{
            $v_s\leftarrow selectVertex(V,M_{sp})$\;
            $V_S\leftarrow V_S \cup \{v_s\}$\;
            $M_{sp}\leftarrow (M_{sp}|B\!F\!S(\{v_s\}, V))$ \;
        }
        $V_S\leftarrow V_S - V_{av}$\;
        \Return{$V_S$}\;
 \caption{Select spread out vertices in a graph}
  \label{algo:SpreadNodes}
\end{algorithm}


Once the number of workers per community has been determined, those numbers must be selected from each community's subgraph (see algorithm \ref{algo:SpreadNodes}).
Finding the optimal solution to maximize the distances between worker nodes is time-exponential in the number of nodes.
We approximate this solution by iteratively adding nodes to the output set, using breadth first searches and maximizing different selection criteria. Time-complexity achieves $O(n_i^2*n + n_i*e)$ this way, with $n$ nodes and $e$ edges in the community $c_i$'s subgraph, $n_i$ the number of workers to place in $c_i$.

\subsubsection{Distancing Criteria}
The main criterion for worker selection is maximizing the distance with the nearest worker. In case of multiple choices, a secondary criterion is used to decide:
(MSR) Maximizing the Sum of a Reward function $r:d \rightarrow reward$:
            \begin{center}
                $r:d \rightarrow \begin{cases}
                0, \qquad d \in \llbracket 0;1 \rrbracket\\
                1, \qquad d=2 \\
                2, \quad\;\: d \in \llbracket 3;+\infty \llbracket\\
                \end{cases}$
            \end{center}
MSR's heuristic takes into account the fact that the primary criterion (maximizing the distance to the closest worker) has operated a pre-selection on the candidate nodes. After the primary criterion, we know that there will be some workers who are close to the candidate. Knowing that, what we seek is to have the highest number of faraway workers (distances of 3 or more). Which is why workers distant by 3 or more yield the reward of 2, while those distant by 2 yield half, and adjacent workers give no reward.

Random choice of the next worker among candidate nodes also gives good results in spreading workers on the graph. However, we require using deterministic methods for verifiability purposes.

\section{Experimental Protocol}\label{experimental protocol}

We will now present the experimental protocol which we will use to evaluate the implementation of our worker selection methods: verifiable random worker selection (\autoref{Vote order-based}) and community detection-based worker selection (\autoref{ComDetWorkerSelection}). The objective of this protocol is to build an environment to quantify and analyze collusion resistance in the protocols that we proposed.

\subsection{Context}

Given an execution of either method placing $n$ workers on a social graph, the probability of collusion will be quantified by the size of the largest clique of workers, distant from each other by at most $k\geq 1$. This metric is time-exponential on $n$, which means experiments on high $n$ values will have to be done with fewer protocol executions. We assume for our experiments that this largest clique of workers of size $m$ will try to collude. We could have chosen to tag a subset of graph nodes as malicious, then measure the size of the largest clique of malicious workers. However, this second technique requires additional assumptions on how malicious workers are distributed on social graphs, which we have chosen to avoid.

We have defined in \autoref{problemstatement} the threshold $t_{max}(n)=\lfloor(1+\sqrt{5+4n})/2\rfloor$ under which a number of $m$ colluding malicious workers still upholds our protocol's security properties.

Given this metric $m$ and $n$ workers placed on a graph, we consider that collusion is possible if $m\geq t_{max}(n)$. Given a set of protocol executions selecting a number of $n$ workers, we also compute a confidence interval $[0; m_{max}]$ on the distribution of obtained $m$ values. A protocol is defined as collusion-resistant with a confidence $c$ for a given number of workers $n$, if the probability that $t_{max}(n) \notin [0; m_{max}]$ is $c$.

\medskip

Our experimental protocol aims to answer the following question: does random sampling of workers among participants give a high enough confidence that no collusion will be possible between those workers? 

\medskip

We will fix the confidence level for our experiments to 95\%. An alternative would have been to compute a p-value of each set of executions for $n$ selected workers. In our case, this p-value would correspond to the probability that we erroneously consider that there is a collusion attack, meaning the probability that no collusion takes place. A high p-value would mean a low probability of collusion. However, due to the reduced number of executions for high $n$ values, this p-value would not be significant, given the low number of considered experiments.



\begin{figure}
 \includegraphics[width=\linewidth, keepaspectratio]{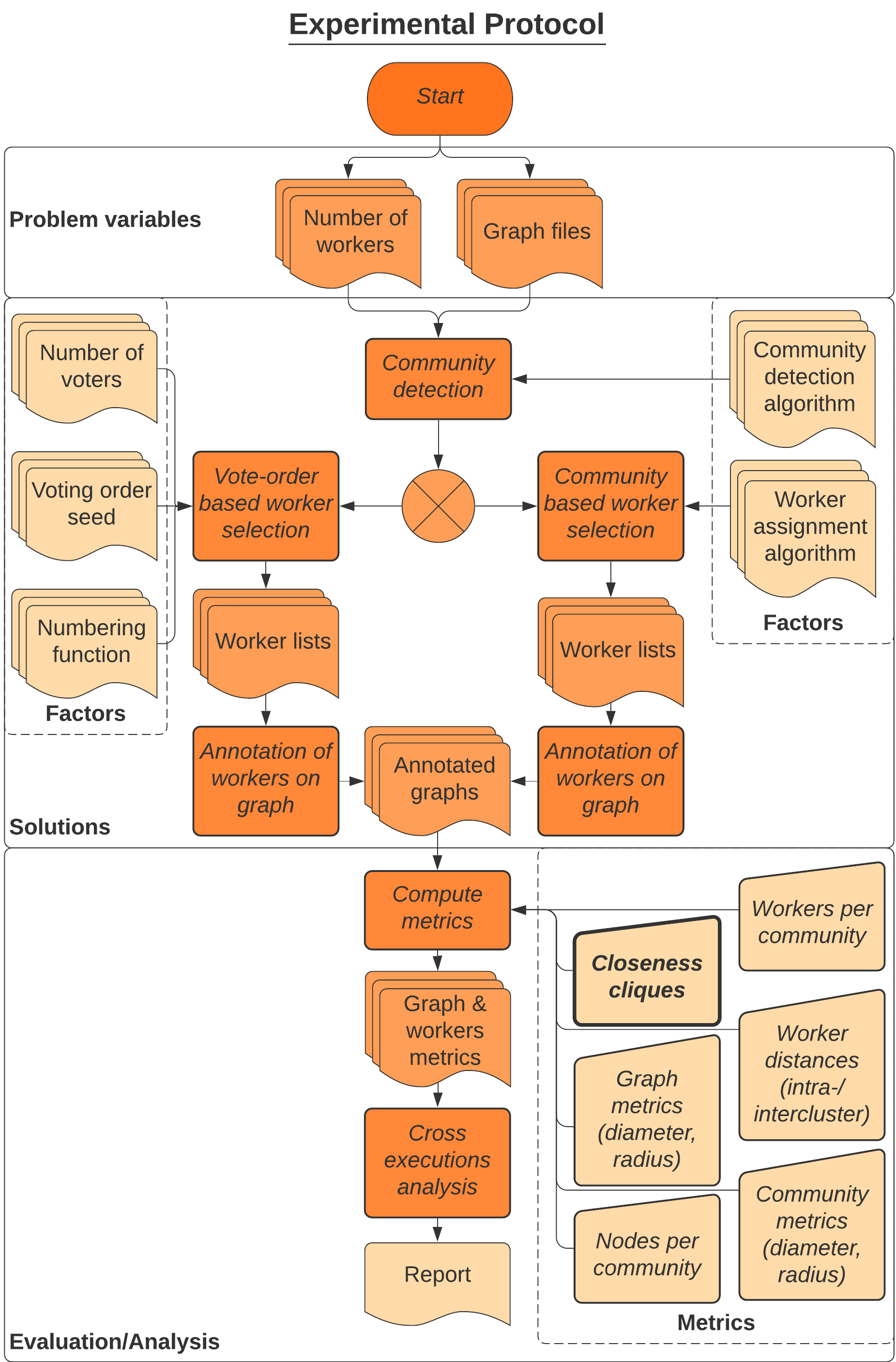}
 \caption{Process diagram for the experimental protocol, given a set of problem variables comprising social graphs and amounts of workers to place. \\Experimentation is divided into a solutions-processing phase followed by metrics computation and evaluation. On the sides of the solution phase, factors for a given algorithm are presented. Our main metric is emphasized in bold.\\This process is fully automated until ''cross executions analysis'' excluded.
}
\label{fig:expProtocol}
\end{figure}

\autoref{fig:expProtocol} describes our experimental protocol's workflow.

The effective parameters which are the number of workers to select were chosen based on 10 points of an exponential growth from 10 workers to $\frac{1}{5}$ of the size of the graph (in the case of facebook\_combined\footnote{See \autoref{dataset_tools} for the dataset introduction}: 807 out of 4039 nodes) with the logspace method from the Python NumPy library\footnote{Python NumPy logspace function documentation at https://numpy.org/doc/stable/reference/generated/numpy.logspace.html}. These parameters will be tested on 100 executions. Due to the time-exponential complexity on the number of workers, experiments for $n \in \{ 1000+200*k | k \in \llbracket 0; 10\rrbracket\}$ will be run on 5 executions.
\begin{table*}[ht]
        \caption{Problem instances or Experimentation inputs}
        \centering
        \begin{tabular}{lc}
            \hline\noalign{\smallskip}
            \emph{Problem Variables} & \emph{Values} \\
            \noalign{\smallskip}\hline\noalign{\smallskip}
            $n_{workers}$ & $\{10, 16, 43, 70, 114, 186, 304, 495, 807\}$\\
            &$\bigcup \{ 1000+200*k | k \in \llbracket 0; 10\rrbracket\}$\\
            \noalign{\smallskip}\hline
        \end{tabular}
        \label{tab:InputPoints}
\end{table*}

\subsection{Randomly Sampling Nodes on a Graph: Verifiable Random Worker Selection Simulation}

We substitute the Verifiable Random Worker Selection with a simulation protocol for the experiment's purpose. This does not have an impact on the results since the simulation keeps the same concept of randomness and the output remains an ordered list of participant IDs.

In lieu of this specific approach, the more general problem of random sampling of workers in a given graph will be evaluated. This gives us an expectable behavior and feasibility approximation for our random number generator protocol.
Those solutions being non-deterministic, they will be considered feasible in practice for a given graph and a number of workers $n$, if the executions’ largest clique-size 95\%-confidence intervals satisfy the problem’s constraints.

In practical terms, this considers a scenario where all $P$ participants are potentially malicious, but a subset of workers can only successfully collude together if they form a clique larger than the security threshold $t$ in the graph.

\subsection{Graph \& worker selection metrics} \label{SelectionMetrics}
Our main metric is the size of the largest cliques of workers distant of at most $k$, $k$ in $\llbracket1;3\rrbracket$. It gets computationally too expensive for $k>3$, for less added insight, as 93\% of communities found in the graph have diameters of 5 or less for the multilevel algorithm, the facebook\_combined graph itself being of diameter 8. The best method should minimize the size of the largest clique given $n_{workers}$.

For behavior analysis purposes, additional metrics will be measured:
\begin{itemize}
        \item Community subgraph diameter and radius;
        \item Number of nodes in communities;
        \item Number of workers in communities;
        \item Distances between workers (intra- and inter-community), mainly on inter-worker distances less or equal to 2. The best method should minimize the amount of workers directly adjacent or separated by one other participant.
\end{itemize}

\section{Chosen Dataset and Tools } \label{dataset_tools}

\subsection{Dataset}

We used the facebook\_combined \cite{fbcombined} dataset for the experimental evaluation. This dataset from Facebook’s social network graph is available through the Stanford network datasets \cite{stanforddatasets}. Some of the properties of the facebook\_combined dataset are listed below.

\begin{itemize}
        \item Undirected edges (symmetrical relationships).
        \item Comprises 4,039 nodes and 88,234 edges.
        \item ``Community detection friendly'': most clusters should be detectable by humans on the graph. Most members of these clusters should be adjacent to multiple other members of the same cluster. Counter-example: streaming website Twitch’s network \cite{twitchdataset}, where some individuals have very high degrees, while their adjacent nodes are mostly not connected to each other.
\end{itemize}

\subsection{Tools}

We used the tools listed below for the experimental evaluation.

\begin{itemize}
\item Graphing tools: igraph (data structures and graph  methods) \cite{igraph}; graph-tool (visualization) \cite{graphtool}.
   
\item Exploration and experimentation: Python Jupyter \cite{jupyter}.

\item Workflow and reproducibility management: A framework developed by Matthieu Bettinger \cite{workflowmanager} (provided with the rest of the source code \cite{sourcecode}); Python Jupyter \cite{jupyter}.

\end{itemize}

\section{Evaluation and Analysis} \label{evaluation}
This part aims to present and analyze the results from the execution of the workflow presented in \autoref{fig:expProtocol}, i.e. the step ``cross-executions analysis'' in the process diagram.

\subsection{Random Nodes Sampling on a Graph}

\begin{figure}
 \centering
 \includegraphics[width=.9\linewidth, keepaspectratio]{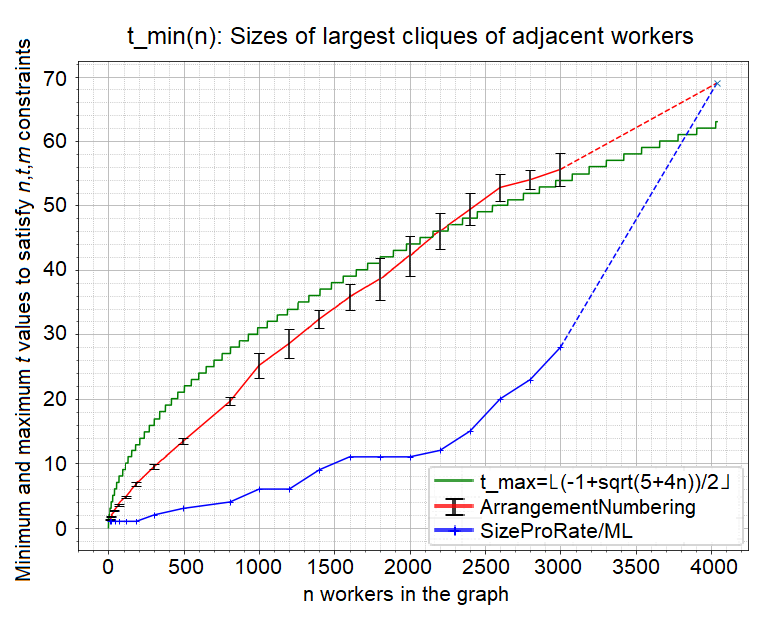}
 \caption{Lower and upper bounds for $t$ in t-n Shamir Secret Sharing Scheme\\
  $t_{max}(n)$ is determined with the problem's strictest constraint on $t$ (\autoref{eqConstr3}, see also \autoref{problemstatement}).
 }
 \label{fig:tBounds1}
\end{figure}

\begin{figure*}
\centering
\begin{subfigure}{\linewidth}
\includegraphics[width=\linewidth, keepaspectratio]{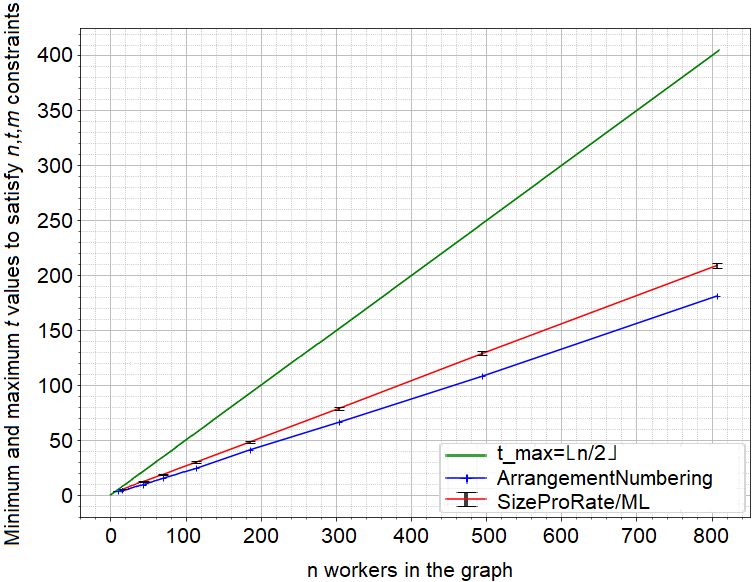}
\label{sfig:atmost2}
\caption{Size of largest cliques of workers distant of at most 2}
\end{subfigure}
 \begin{subfigure}{\linewidth}
 \includegraphics[width=\linewidth, keepaspectratio]{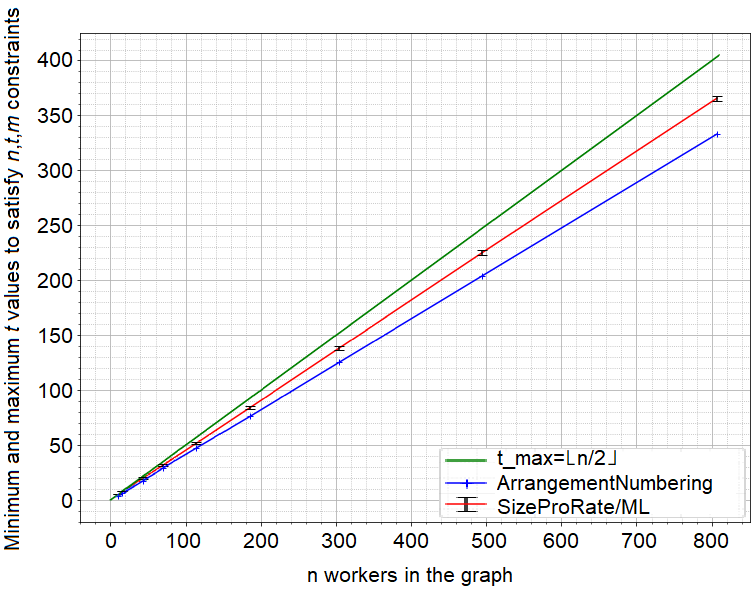}
 \label{sfig:atmost3}
 \caption{Size of largest cliques of workers distant of at most 3}
\end{subfigure}

 \caption{Lower and upper bounds for $t$ in t-n Shamir Secret Sharing Scheme\\
  $t_{max}(n)$ is obtained when removing the problem's strictest constraint on $t$ (\autoref{eqConstr3}, see also \autoref{problemstatement}).
 }
 \label{fig:tBounds2}
\end{figure*}
 
We plot the lower and upper bounds for $t$ in the $t-n$ Shamir Secret Sharing Scheme in Figures \ref{fig:tBounds1} and \ref{fig:tBounds2}. The inequalities \ref{eqConstr1} through \ref{eqConstr3} from \autoref{problemstatement} give the maximum values $t_{max}$ given $n$. Meanwhile, the metric we evaluate empirically, that is the size of the largest clique of workers distant of at most $k$ (with $k$ increasing from 1 to 3 in Figures \ref{fig:tBounds1}, \ref{fig:tBounds2}.a and \ref{fig:tBounds2}.b), gives the lower bound $t_{min}$. This is done for our $ArrangementNumbering$ and $SizeProRata/ML$ (Multi-Level) worker selection algorithms, presented respectively in sections \ref{Vote order-based} and \ref{ComDetWorkerSelection}. A solution is viable for a given $n$ if $t_{max}>t_{min}$. 
$ArrangementNumbering$'s curve is given with a 95\% confidence interval (error bars). For $n\leq807$, 100 executions were done for each point on $ArrangementNumbering$, 5 executions for $n>807$, due to the metric's exponential complexity on $n$. No error bars are necessary for $SizeProRata/ML$, as the algorithm is deterministic.

Intuitively, a method which has access to the information upon which a metric is computed will perform better on that metric than a method without that information. In our case, the metric being the size of the largest clique of workers, our graph-aware proposed solution should perform better than our random worker selection protocol.
This observation can be confirmed by the graph representing the cliques' size threshold for the $t-n$ Shamir Secret Sharing Scheme (see Figure \ref{fig:tBounds1}). Undeniably, the $SizeProRata/ML$ curve (``ML'' stands for multi-level community detection), our graph-aware solution, shows considerably better results than the random selection with $ArrangementNumbering$.
Until 45\% of participants are workers, $ArrangementNumbering$ is closely under the maximal threshold curve : there is a confidence of 97.5\% that the size of largest cliques is under $t_{max}$ until ~1800 workers.
$SizeProRata/ML$ is remarkably more efficient and its heuristics only begin to lose efficiency at around half of the voters as workers.
When the number of workers reaches the number of voters, the $ArrangementNumbering$ and $SizeProRata/ML$ curves will intersect in a final point. This intersection corresponds to the largest clique size (in this case 69 for 4039 nodes) when all the voters are workers. Computations for over 3000 workers were not executed, because of the largest clique search algorithm’s time-complexity. Only the final intersection point was computed. Dotted lines show the expected evolution of both curves between 3000 workers and the endpoint.

However, considering a worker threshold of $\frac{1}{5}$ of the participants (here 807 among 4039), which is already a high number of workers, the two workers selection methods stay under the $t_{max}$ curve determined by our problem’s constraints.
\subsection{Results under relaxed constraints}
As defined in \autoref{problemstatement}, our main constraint on the upper bound of the security threshold $t$ is due to the checksum verification which implies $n-t^2 > m$ (Inequality \ref{eqConstr3}). Should the constraint of Inequality \ref{eqConstr3} be removed or be absent (i.e. in another protocol without this checksum mechanism), only constraints linked to Shamir’s Secret sharing scheme would remain (Inequalities \ref{eqConstr1} and \ref{eqConstr2}, namely $t>m$ and $n-t>m$), which give an upper bound for t: $t_{max}=\lfloor \frac{n}{2}\rfloor$. In practical terms, as $n-t\geq n-t^2>m$ for $t\in \mathbb{N}^*$, removing Inequality \ref{eqConstr3} enables tolerating a greater proportion of malicious workers, for a given number $n$ of chosen workers.

If we were to consider this new upper bound, we could compare curve behaviors of both our selection methods with that new boundary curve, for cliques of workers distant of at most 2 (see Figure \ref{fig:tBounds2}.a), resp. 3 (see Figure \ref{fig:tBounds2}.b). These methods’ curves greatly surpass the more constrained version of $t_{max}$.
In case of a maximal distance of 2 resp. 3, we can see that both methods representing $t_{min}$ grow linearly with a slighter slope than the boundary curve representing $t_{max}$ ($SizeProRata/ML$: 0.23 resp. 0.41; $ArrangementNumbering$: 0.26 resp. 0.43; $t_{max}(n)$: 0.5). Therefore, should the $n-t^2>m$ constraint be lifted whilst maintaining the protocol’s properties, better collusion-resistance insights can be obtained. Indeed, we would know that under these new constraints, it is unlikely that a clique of workers distant of at most 3 would be larger than $t_{max}$.

\section{Discussion} \label{discussion}




\subsection{Clique Sizes in Social Graphs}
When a social graph increases in size, how do graph properties like the size and the number of the largest cliques scale?

The denser a given graph, the larger the probable size of the largest clique and the number of cliques of a given size. If the proportion of nodes in big cliques compared to the global graph gets higher, random sampling of nodes will more and more likely occur in those cliques. The same reasoning holds true for big communities, a relaxed concept of cliques of nodes. Johan Ugander et al. described the structure of Facebook's social graph in 2011\cite{ugander2011anatomy}. For both the U.S. and global friends networks on Facebook, around 90\% of users had less than 500 friends and 1\% had more than a thousand friends (maximum number at around 5000 friends). They used the degeneracy metric in their analyses, which corresponds in ''an undirected graph $G$ [to] the largest $k$ for which $G$ has a non-empty $k$-core. Meanwhile, the $k$-core of a graph $G$ is the maximal subgraph of $G$ in which all vertices have degree at least $k$''. A $k$-core corresponds to a $k+1$-clique if its size is $k+1$. This also means that $k$-cores are sets of nodes which may contain cliques of size less or equal to $k+1$. Therefore, degeneracy provides an upper bound for the size of the largest clique in the graph. Their findings use the degeneracy for nodes of a certain degree, i.e. for a given person, the maximum $k$ friends which also know $k-1$ other friends of that same person. Degeneracy grew monotonously with the node degree. The maximal degeneracy for the $95^{th}$ percentile was of around 200 for a degree of 5000. This would mean that 200 friends of that person knew 199 other friends of that person. What interests us more is that a clique containing that person would be of size lower than 200. Their analysis was carried out on active Facebook users (users, with at least one friend, who logged in at least once in the last month prior to the analysis), representing a graph with 721 million nodes. This result on such a large graph gives us reassurance on the evolution of the size of the largest clique as the graph gets larger. Indeed, for such a graph size, $t_{max}=26 851$, which is two orders of magnitude bigger than the upper bound for the largest clique's size. In fact, a graph of size 39,799 (10x the size of the graph we used) would still tolerate a largest clique of 200 nodes whilst ensuring collusion-resistance.

\subsection{On the Feasibility of Random Worker Selection}
A determining factor on whether random selection of workers is feasible with a low collusion probability in a given graph is the size of the graph's largest clique (see Figure \ref{fig:tBounds1} for reference). If that size is significantly lower than the upper bound for the $t-n$ Shamir Secret Sharing Scheme's $t$ value obtained through the problem's constraints, then there is a low collusion probability, i.e. high collusion resistance, for any number of workers. However, if the size of the largest clique is close to or greater than $t_{max}(P)$, with $P$ the total number of nodes in the graph, then there exists an upper bound $n_{max}$ for $n$ where, for $n\geq n_{max}$, $P(|Clq_{max}|>t_{max})>5\%$, with $Clq_{max}$ the largest clique's set of nodes for $n$ workers.

If the graph is unknown or unavailable, then this upper bound $n_{max}$ can be approximated through other approaches. Insights about the voting population's social structure, for example insights on its density, the (expected or known) size of some communities among participants, can help in estimating $n_{max}$. Without such graph knowledge, then one can use system constraints (Inequalities \ref{eqConstr1}-c) and the hypergeometric distribution followed by this random selection (as presented in \autoref{Vote order-based}). By estimating a certain proportion of $M$ malicious participants among the total $P$, one can get all numbers of workers $n$ such that system constraints are verified.

\subsection{On the Feasibility of Relaxed Graph-Aware Methods}

Would less informed knowledge about the graph be sufficient in order to give high collusion-resistance confidence?

For low values of $n_{workers}/n_{participants}$, the size of the largest clique of workers seems to increase linearly with the number of workers (see Figure \ref{fig:tBounds1}). That function's slope is initially steeper than the one of the linear function passing through the point corresponding to the largest clique in the graph, then the slope gets slighter in order to end on that same point. It would be of interest to investigate if that observation still holds on other social graphs. In that case, because $t_{max}$ follows a square root-shaped function, there exists a range of low values of $n$, where $t_{max}$ is greater than $t_{min}$ (size of largest clique with 95\% confidence).

Other criteria altogether, not using a graph, could be used to help determine a number of workers $n$:
\begin{itemize}
        \item Using the curve obtained by intersecting the hypergeometric law's Cumulative Distribution Function $P(n,m)$ with a plane of probability $p$ (e.g. 95\%) as defined in \autoref{Vote order-based properties}. This method requires quantifying an expected total number of malicious participants $M$;

        \item Optimizing $n$ on the criterion of the amount of needed messages. Indeed Shamir's Secret Sharing Scheme requires dividing each one of the $P$ participant's vote in $n$ shares, one per worker. This means there will be $n*P$ messages placed on the distributed ledger during the referendum-vote. For scalability purposes, the number of workers should be kept as low as security criteria permit it.
\end{itemize}

\subsection{On the Feasibility of Graph-Aware Methods}
Should a social graph of participants be available for a given referendum, where should it be stored?

If only the referendum initiator has access to it, should they be malicious, nothing prevents them from not using it altogether in designating workers. Forcing him to use it could be done by forcing him to provide a proof that the result was obtained through the algorithm.

Let us now consider a graph annotated with participant IDs on nodes. The initiator can use it or transmit it to other malicious entities to violate participant anonymity, through graph inference re-identification.

However, knowing only the unannotated graph may provide lower and upper bounds for $n$ in the same way our experiments did (see Figure \ref{fig:tBounds1}). Anonymity could be maintained in this case.

If the graph is public, then it becomes easier to ascertain whether the initiator used the algorithm, for example through Smart Contracts (Ethereum)\cite{ethereum}. However, an annotated graph would again be at risk of participant re-identification.
If we were to divide the graph among participants in order to decentralize tasks, we would need workers and would therefore have the same problem to select those.

Another variant would be to use a decentralized social graph, with participants knowing only their ``friends'' on the now implicit social graph (plus some strangers to avoid re-identification if some nodes have a low degree). A decentralized algorithm should then be designed to select workers under those constraints.

\section{Conclusion}
\label{conclusion}

In this paper, we proposed two solutions to provide better collusion-resistance in distributed protocols where a subset of workers needs to be selected from the set of participants. The referendum voting protocol of Schiedermeier et al. \cite{schiedermeier2019transparent} serves as a use-case for our solutions. In this protocol, which is based on blockchain and Secure Multi-Party Computation, a subpopulation of participants is arbitrarily chosen by an initiator entity to compute the referendum's calculations. In this referendum use-case, worker collusion can lead to corruption of the final result. The referendum could also be rendered void by massive malicious worker inactivity. It is desirable to prevent these types of behaviors and to avoid relying on single entities in our use-case but also in the more general context of decentralized or multi-party systems.

In order to prevent collusion, we introduced two worker selection protocols: a verifiable random worker selection based on decentralized computation of a random seed, as well as a selection based on community detection in social graphs.
Firstly, we used the blockchain's immutability and ordering to design a collusion-resistant decentralized protocol to randomly select workers.
Secondly, we considered a social graph representing participants and proposed an algorithm to distance workers from each other in the graph.

Based on a social graph and our problem's constraints, we computed the size of the largest clique of workers to evaluate the number of workers' bounds for which our solutions were resistant to collusion with high confidence.

Both approaches provided ranges of numbers of workers satisfying the constraints (see \autoref{fig:tBounds1}). The decentralized random worker selection works from low numbers of workers to an upper limit which depends on the size of the graph's largest clique. As expected, the method taking advantage of the graph's structure provides better results: it distances workers better for a wider range of numbers of workers.

As discussed, an interesting topic for future work would be to analyze in depth the impact of the social graph structure on the protocol's resistance to collusion. An equally important topic would be to fully decentralize the method based on the graph structure and community detection. This solution should ensure privacy for participants, be transparent to all and verifiable by all, whilst preserving the demonstrated collusion-resistance properties.

\bibliographystyle{spmpsci}          
\bibliography{bibliography}   

%
%

\section*{Declarations}

All manuscripts must contain the following sections under the heading 'Declarations'.

If any of the sections are not relevant to your manuscript, please include the heading and write 'Not applicable' for that section.

To be used for all articles, including articles with biological applications

\subsection*{Funding (information that explains whether and by whom the research was supported)}
Not applicable.
\subsection*{Conflicts of interest/Competing interests (include appropriate disclosures)}
On behalf of all authors, the corresponding author states that there is no conflict of interest.
\subsection*{Availability of data and material (data transparency)}
Dataset: facebook\_combined \cite{fbcombined} dataset available through the Stanford network datasets \cite{stanforddatasets}.

\subsection*{Code availability (software application or custom code)}
Source code used in experiments available on Github \cite{sourcecode}.

\subsection*{Authors' contributions (optional: please review the submission guidelines from the journal whether statements are mandatory)}

\subsection*{Additional declarations for articles in life science journals that report the results of studies involving humans and/or animals}
Not applicable.

\subsection*{Ethics approval (include appropriate approvals or waivers)}
Not applicable.

\subsection*{Consent to participate (include appropriate statements)}
Not applicable.

\subsection*{Consent for publication}
Not applicable.

\end{document}